\documentclass[
  aps,
  pre,
  twocolumn,
  reprint,
  amsfonts,
  amssymb,
  amsmath,
  superscriptaddress,
]{revtex4-2}

\setcitestyle{super}

\usepackage[utf8]{inputenc}
\usepackage{bm}  
\usepackage{siunitx}
\DeclareSIUnit\angstrom{\text{Å}}
\usepackage{booktabs}
\usepackage{graphicx}
\usepackage[svgnames]{xcolor}
\usepackage{microtype}
\usepackage[raggedright, compact]{titlesec}  
\usepackage{comment}
\usepackage{csquotes}
\usepackage[normalem]{ulem}  
\usepackage{ifthen}

\usepackage{ragged2e}
\usepackage{caption}
\DeclareCaptionJustification{reallyjustified}{\justifying}  
\titleformat*{\subsection}{\bfseries\large\sffamily}
\titlespacing*{\subsection}{0pt}{*3}{0pt}

\titleformat*{\subsubsection}{\bfseries\normalsize\sffamily}
\titlespacing*{\subsubsection}{0pt}{*3}{0pt}

\titleformat*{\paragraph}{\bfseries\small}
\titlespacing*{\paragraph}{0pt}{*1}{*2}  

\titleformat*{\subparagraph}{\bfseries\footnotesize\sffamily}
\titlespacing*{\subparagraph}{0pt}{*1}{*2}  

\usepackage[hidelinks]{hyperref}
\usepackage{enumitem}

\newcommand\wordcount{
    \immediate\write18{texcount -sub=section \jobname.tex  | grep "Section" | sed -e 's/+.*//' | sed -n \thesection p > 'count.txt'}
(\input{count.txt}words)}

\newcommand*{\vb}[1]{\mathbf{#1}}   

\newcommand*{\vvec}{\vb{v}}

\newcommand{\bolds}{\mathbf {s}}
\newcommand{\boldv}{\mathbf {v}}

\newboolean{ColourText}
\setboolean{ColourText}{true}

\def\red#1{\textcolor{black}{#1}}

\ifthenelse{\boolean{ColourText}}{
  \newcommand*{\ADD}[1]{{\color{black} #1}}
  \newcommand*{\DEL}[1]{\textcolor{MediumPurple}{\sout{#1}}}

}{
  \newcommand*{\ADD}[1]{#1}
  \newcommand*{\DEL}[1]{}

}

\begin{document}

\title{Quantum vortices leave a macroscopic signature 
in the thermal background}

\newcommand*{\Lagrange}{%
  Université Côte d'Azur, Observatoire de la Côte d'Azur, CNRS,
  Laboratoire J.~L.~Lagrange, Boulevard de l'Observatoire CS 34229 - F 06304 NICE Cedex 4, France
}

\newcommand*{\NCL}{School of Mathematics, Statistics and Physics, 
Newcastle University, Newcastle upon Tyne NE1 7RU, United Kingdom}

\newcommand*{\CNR}{Istituto per le Applicazioni del Calcolo M. Picone,
IAC-CNR, Via dei Taurini 19, Roma 00185, Italy}

\author{Luca Galantucci}
\email{luca.galantucci@cnr.it}
\affiliation{\CNR}
\affiliation{\NCL}

\author{Giorgio Krstulovic}
\email{giorgio.krstulovic@oca.eu}
\affiliation{\Lagrange}

\author{Carlo F. Barenghi}
\email{carlo.barenghi@newcastle.ac.uk}
\affiliation{\NCL}

\date{\today}


\begin{abstract}
Recent work has highlighted the remarkable properties of quantum
turbulence in superfluid helium~II, consisting
of a disordered tangle of
quantised vortex lines which interact with each other
and reconnect when they collide. According to Landau's
two-fluid theory, these vortex lines move
in a surrounding of thermal excitations called the {\it normal fluid}.
Until now, the normal fluid has often
been considered a passive background which simply provides the
vortex lines with a mechanism for dissipating their kinetic
energy into heat. 
Using a model which fully takes into account the two-way
interaction between the vortex lines and the normal fluid, here
we show \ADD{numerically} that each vortex line creates a macroscopic wake in the
normal fluid that can be larger than the average 
distance between vortex lines; this is surprising, given the microscopic
size of the superfluid vortex cores which induce these wakes.
We show that in heat transfer experiments, 
the flow of the normal fluid can therefore
be described as the superposition of an imposed uniform flow 
and wakes generated by the vortex lines, leading to non-classical 
statistics of the normal fluid velocity. We also argue that this
first evidence of independent fluid structures
in the thermal excitations postulated by Landau may be at the root 
of recent, unaccounted for, experimental findings.
\end{abstract}

\maketitle

It was Landau who first understood that the properties of a many-body
quantum system such as helium~II depend on thermally excited
elementary excitations - collective modes which he 
called phonons and rotons depending on
whether their dispersion relation is linear or quadratic.
This idea has influenced many areas of modern physics since. 
In the context of liquid helium, Landau's intuition led to 
the formulation of the two-fluid theory, which 
accounts for unusual behaviours of helium~II,
from second sound and thermal counterflows \cite{vinen-1957c} to 
critical velocities \cite{babuin-stammeier-varga-rotter-skrbek-2012}.  
Briefly, helium~II behaves as the mixture of two inseparable fluid components:
the viscous normal fluid (associated to the thermal
excitations) and the inviscid superfluid (associated to the ground state). 
Each component has its own density field and velocity field,
$\rho_s$, $\vvec_s$ for the superfluid and $\rho_n$, $\vvec_n$
for the normal fluid, where $\rho=\rho_n+\rho_s$ is the total density.
As a consequence, two distinct motions occur simultaneously at the
same position in space.
Of the two components, only the normal fluid has nonzero entropy 
(hence is capable of carrying heat). Because of quantum mechanical
constraints, the superfluid's vorticity is concentrated to vortex lines
of quantised circulation (in units of $h/m$ where $h$ is Planck's
constant and $m$ is the mass of one helium atom) and microscopic thickness 
(the vortex core radius is only $a_0 \approx 0.1~\rm nm$). 
Over the years, the vortex lines have 
attracted great attention, revealing remarkable properties: 
vortex lattices \cite{Yarmchuck-etal-PRL-1979},
vortex nucleation \cite{Avenel-Varoquaux-PRL-1985},
Kelvin waves \cite{Makinen-etal-NatPhys-2023,peretti-etal-2023},
vortex reconnections \cite{Fonda2014}, and
quantum turbulence \cite{Skrbek-etal-PNAS-2021} 
of various kinds \cite{barenghi-etal-AVS-2023}. 

In contrast, much less attention has been paid to 
the normal fluid, of which we still know very little. 
We know that Landau's excitations are scattered by the velocity
field of the vortex lines, creating a mutual friction 
between the two fluid components which allows energy exchange \cite{hall-vinen-1956a,hall-vinen-1956b}. 
For the sake of
simplicity, in most problems it is usually assumed that the normal fluid
is spatially uniform or at rest, simply providing a dissipative background
to the vortex lines. 

Based on the mean
free path $\lambda$ and the Knudsen number $\lambda/d$,
the description of Landau's excitations as a fluid (rather than a ballistic
gas) is appropriate for lengthscales
$d > 0.2~\mu{\rm m}$ at $T=1.5~\rm K$ and $d>0.02~\mu{\rm m}$ 
at $T=2.1~\rm K$, which is the temperature range of interest here.
These values of $d$ are safely smaller than both the size
of tracer particles used to visualize flows ($\gtrsim 1\mu{\rm m}$)
\cite{zhang-vansciver-2005,bewley-lathrop-sreenivasan-2006,lamantia-duda-rotter-skrbek-2013d,tang-etal-2023,kubo-tsuji-2019} 
and the typical distance between vortex lines in turbulence experiments, 
$\ell \approx 10$ to $100~\mu{\rm m}$.
It is therefore natural to assume that the normal 
fluid obeys no-slip boundary conditions, but what should be the normal
fluid's large-scale velocity profile in a channel is still an open question.
For coflows ($\vvec_s$ and $\vvec_n$ flowing in the
same direction, e.g. driven by bellows) a laminar normal fluid 
profile has been observed \cite{xu-vansciver-2007} but its scaling
with the Reynolds number has not been explained yet.
For counterflows ($\vvec_s$ and $\vvec_n$ flowing in opposite direction 
driven by a heater) there is theoretical debate about the
shape of the laminar profile
\cite{galantucci-sciacca-barenghi-2015,yui-etal-2018},
and experimental evidence (but no quantitative information) 
of a transition to turbulence at large velocity 
\cite{melotte-barenghi-1998,Guo-etal-PRL-2010}. 
Until now, no other flow structure, at any scale, has ever been identified in the
normal fluid, which has remained somewhat mysterious.

\begin{figure*}[htbp]
  \centering
  \includegraphics[width=.98\textwidth]{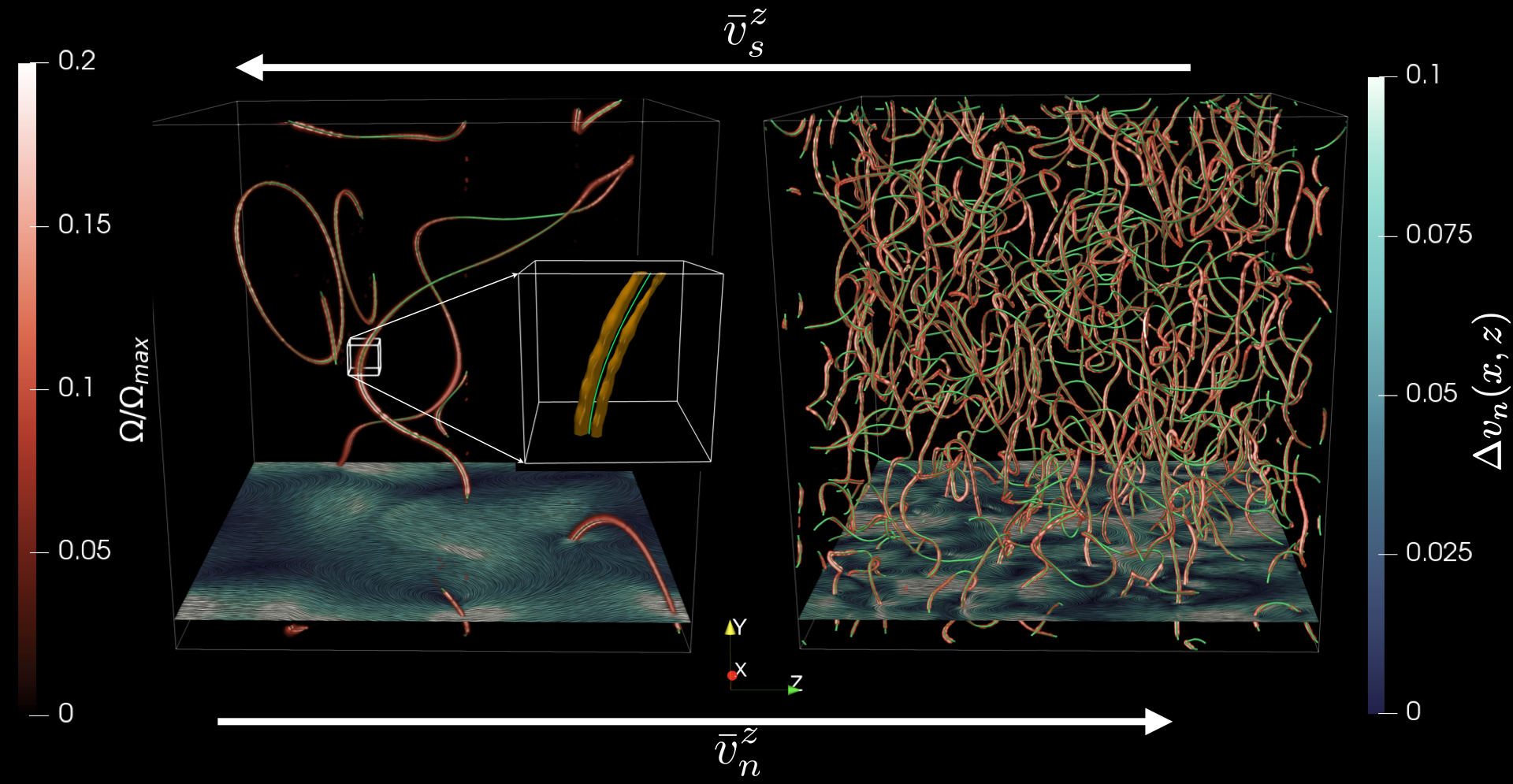}
\caption{
{\bf Snapshots of vortex tangles.} Turbulence at $T=1.5~\rm K$ generated by
counterflow velocities $v_{ns}^{(1)}= 0.27 \rm cm/s$ (left) and
$v_{ns}^{(2)}= 0.94 \rm cm/s$ (right).
The superfluid vortex lines are displayed as green curves and the
normal fluid dipoles are visualized by the normalised enstrophy
$\Omega(\mathbf{x})/\Omega_{max}$ in reddish color.  
The relative magnitude of normal fluid fluctuations $\Delta v_n (x,z)$
is plotted on a $xz$ plane at constant $y$.}
\label{fig1} 
\end{figure*}

The numerical results that we present here predict that the relative
motion of vortex lines and the normal fluid creates, besides small
dipolar structures\cite{kivotides-barenghi-samuels-2000}
 which are probably too small to be observed, 
also large macroscopic wakes in the normal fluid. 
The wakes are so large (even larger than the typical inter-vortex 
distance $\ell$)
that we \ADD{infer from our data that} they affect the velocity statistics of the normal fluid, 
producing skewed distributions
with wide tails, strikingly dissimilar from classical turbulence.
We shall see that some indirect evidence of this effect has been measured
in counterflow turbulence experiments without being recognized.

Our numerical simulations of counterflow turbulence use
the model FOUCAULT \cite{galantucci-krstulovic-etal-2020}. The code models
vortex lines as space-curves of infinitesimal thickness, which is
appropriate given the huge separation of scales between the vortex core
thickness $a_0$ and the typical distance $\ell$ between vortex lines in quantum turbulent
flows. Unlike the original one-way
approach of Schwarz \cite{schwarz-1988}, FOUCAULT accounts for the two-way 
interaction between the vortex lines (which evolve according to the
Biot-Savart law and mutual friction corrections) and the normal fluid (which evolves according to a modified
Navier-Stokes equation) \cite{kivotides-2018}; as in the approach of Schwarz,
vortex reconnections are implemented algorithmically.
For simplicity our calculations are performed in a cube of 
size $D=0.1 \rm cm$ with periodic boundary conditions (see Methods for details).

To model thermal counterflow, we impose average normal fluid and superfluid 
velocities  along the positive and negative $z$-directions, respectively. 
This creates a counterflow velocity 
$v_{ns}=\vert \bar{v}_n^z - \bar{v}_s^z \vert$ 
(where overbars denote spatial averages) which, in the 
experiments, is proportional to the applied heat flux. The two fluid components
flow in opposite directions: the normal fluid 
carries the heat away from the heater at speed $\bar{v}_n^z=(\rho_s/\rho) v_{ns}$, 
while the superfluid flows in the opposite direction to conserve mass, $\bar{v}_s^z=-(\rho_n/\rho) v_{ns}$. 
The initial condition consists of few random vortex rings, which, after 
a transient, evolve into a statistically steady state, which is independent 
of the initial condition; at this point the vortex line density 
(defined as the length of vortex lines per unit volume) fluctuates around
a mean value $L$ corresponding to the average intervortex distance
$\ell \approx 1/\sqrt{L}$. 
Our numerical experiments are performed at temperature $T=1.5~\rm K$ at
two distinct values of the counterflow velocity: 
$v_{ns}^{(1)}= 0.27 \rm cm/s$ and
$v_{ns}^{(2)}= 0.94 \rm cm/s$, yielding $L^{(1)}=7.2 \times 10^2 \rm cm^{-2}$ 
and $L^{(2)}=1.1 \times 10^4 \rm cm^{-2}$ respectively.
%
These turbulent vortex tangles are anisotropic, as vortices tend
to lie on $xy$ planes perpendicular to the applied counterflow in
the $z$ direction.

\begin{figure*}[htbp]
  \centering
  \includegraphics[width=.45\textwidth]{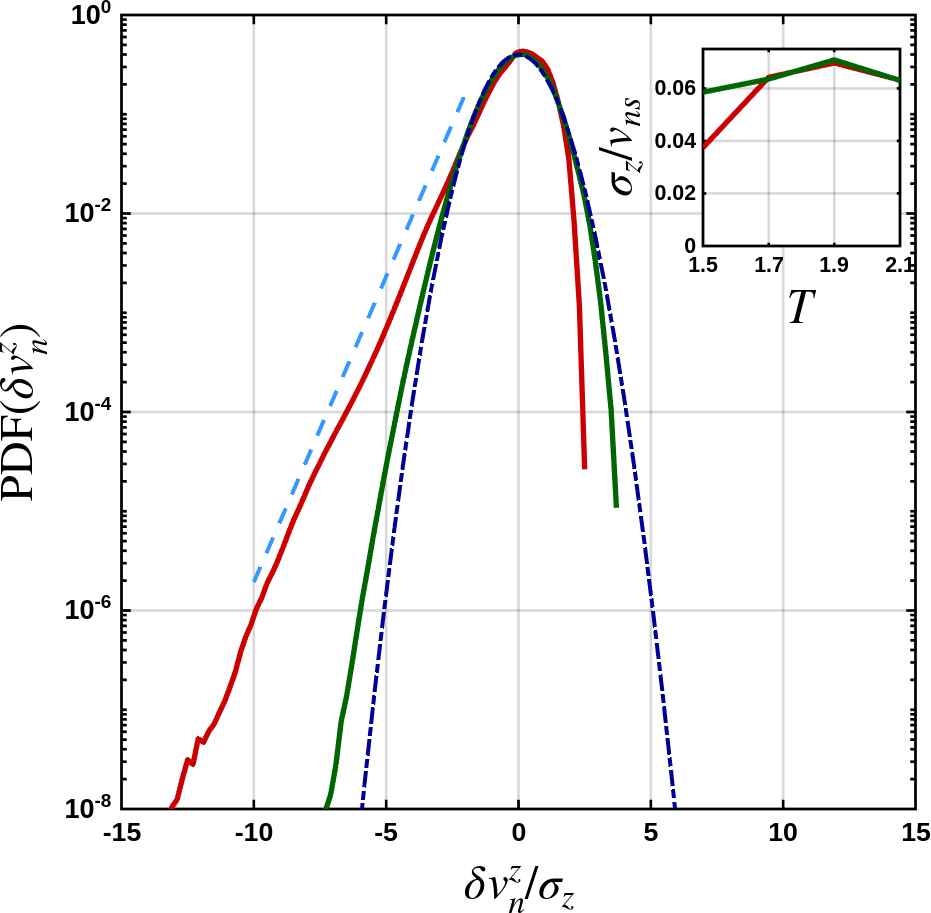}\hspace{0.06\textwidth}
  \includegraphics[width=.45\textwidth]{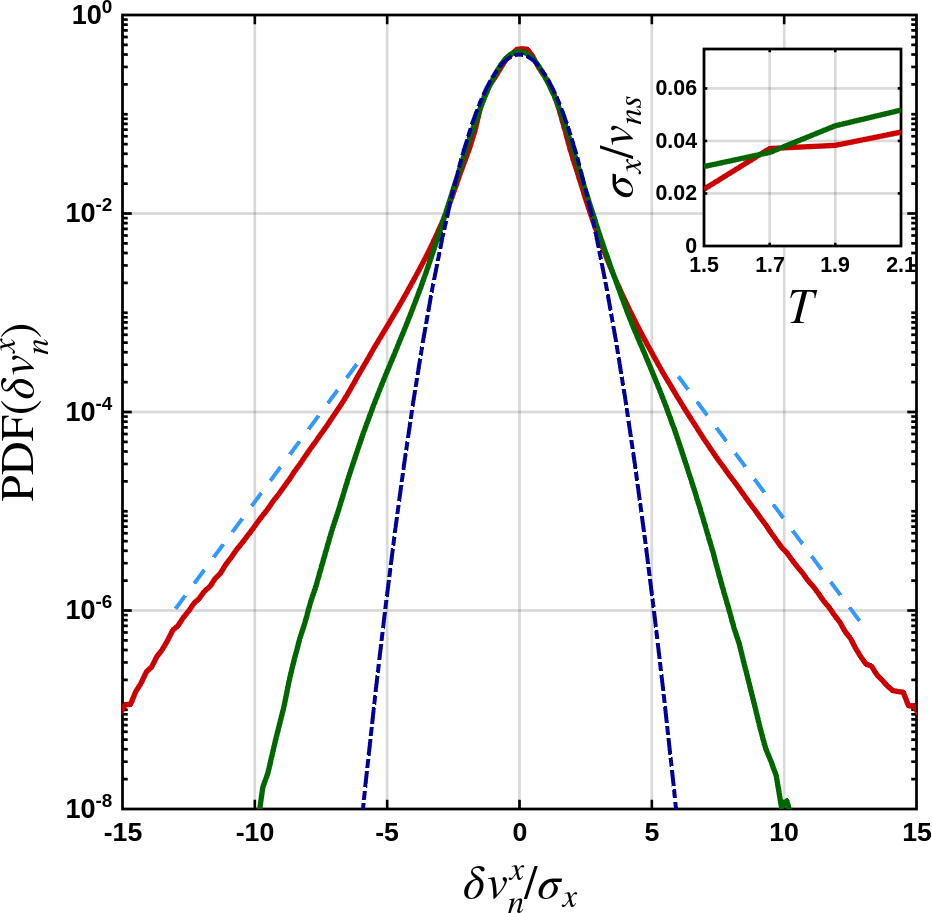}
\caption{
{\bf PDFs of velocity fluctuations.} Calculations at $T=1.5~{\rm K}$
at the same parameters as Fig.~\ref{fig1}.
Left: streamwise PDF$(\delta v_n^z)$ vs $\delta v_n^z/\sigma_z$ 
where $\sigma_z$ is the standard deviation.
Right: spanwise PDF$(\delta v_n^x)$ vs $\delta v_n^x/\sigma_x$
where $\sigma_x$ is the standard deviation.
Red curves refer to $v_{ns}^{(1)}=0.27 \rm cm/s$ and green curves to
$v_{ns}^{(2)}=0.94 \rm cm/s$. Gaussian distributions are showed 
in dot-dashed dark blue line for reference. Dashed cyan curves are
exponential fits to the wide tails.
{\bf Insets.} Left (right): streamwise (spanwise) normalised standard deviation $\sigma_z/v_{ns}$
($\sigma_x/v_{ns}$) as a function of temperature $T$. Colors as in main figure.
}
\label{fig2}
\end{figure*}

Figure~\ref{fig1} shows snapshots of the computed vortex tangles corresponding to 
$v_{ns}=v_{ns}^{(1)}$ (left) and $v_{ns}=v_{ns}^{(2)}$ (right). 
The superfluid vortices are displayed as 
green curves, and the enstrophy of the normal fluid, defined as
 $\Omega(\mathbf{x})=|\bm{\omega}_n(\mathbf{x})|^2/2$ 
(where $\bm{\omega}_n=\nabla \times \vvec_n$) and normalised by
its maximum value $\Omega_{max}$, is rendered in reddish colours 
according to the colormap reported on the left. 
At the bottom 
of each panel, we report a two-dimensional slice on a plane
at fixed $y=y_0$ of the relative magnitude of in-plane normal fluid fluctuations
$\Delta v_n (x,z)=||\delta\tilde{{\bf v}}_n(x,y_0,z)||/|\bar{v}_n^z|$,
with $\delta\tilde{{\bf v}}_n=(v_n^x,v_n^z -\bar{v}_n^z)$, 
colour-coded by the colormap on the right.

Previous calculations
\cite{kivotides-barenghi-samuels-2000,galantucci-krstulovic-etal-2023,
inui-tsubota-PRB-2021} have shown that each superfluid vortex is 
surrounded by two localised regions of large normal fluid enstrophy,
essentially a vorticity dipole induced by the friction. This dipole is small,
of the order of only few micrometers 
($\approx 2$ to $ 4\,\mu m \,$ \cite{galantucci-krstulovic-etal-2023}).
The novel effect which here we report is the existence of much larger
normal fluid structures, whose existence has been to some extent speculated \cite{mastracci-etal-2019}, but not clearly observed. 
The two-dimensional slices for both low (left) and high (right) counterflow velocity regimes reported in Fig.~\ref{fig1}, clearly show that in the proximity of superfluid vortices there are regions (almost two orders of magnitude larger than the dipoles) where 
$\vvec_n$ is significantly different than $\bar{\vvec}_n$
(the magnitude of the normal fluid velocity fluctuations being up to
10$\%$ the applied normal fluid velocity $\bar{v}_n^z$).
This is the first clear evidence of large scale (potentially observable experimentally) fluid structures which spontaneously appear in Landau's sea of thermal excitations.

Such large scale structures strongly modify and impact the statistical distribution of the normal fluid 
velocity.  We concentrate the attention on the streamwise and spanwise velocity fluctuations, 
defined as $\delta {v}_n^z =v_n^z - \bar{v}_n^z$ and $\delta {v}_n^x =v_n^x$ (since $\bar{v}_n^x=0$), 
respectively.  In Fig.~\ref{fig2} we plot the probability distribution functions (PDFs) of the 
velocity fluctuations normalised by their respective variances $\sigma_z$ and $\sigma_x$, for both 
counterflow values.
$v_{ns}^{(1)}=0.27 \rm cm/s$ (red) and $v_{ns}^{(2)}=0.94 \rm cm/s$ (green),
where $\sigma_z$ and $\sigma_x$ are the standard deviations of streamwise and spanwise velocities 
respectively.

%

The main result, shown in the left panel of Fig.~\ref{fig2}, is that the PDFs of the streamwise 
velocity fluctuations are left-skewed, showing the predominance of negative fluctuations. 
In other words there are large regions where the normal fluid flows slower than the 
average flow in the direction parallel to the applied counterflow. 
In the spanwise direction, on the contrary, the PDFs are symmetrical - see
the right panel of Fig.~\ref{fig2}. This effect is independent of the counterflow
velocity; in the low counterflow case (red curves), we observe that the tails have an exponential behaviour.

Additionally, we find that, consistently with Particle Tracking Velocimetry 
(PTV) experiments at not too small counterflow velocities 
\cite{mastracci-guo-2018}, 
the standard deviation of the streamwise velocity is larger than its 
spanwise counterpart: $\sigma_z \approx 2 \sigma_x$, \ADD{as clearly shown in the insets of Fig.\ref{fig2}}. 
As we move from the low to high velocity regime, the tails and the skewness 
of the PDFs become less pronounced, while the standard deviations increase \ADD{($\sigma/v_{ns} \sim \text{constant}$, Fig.\ref{fig2} insets)},
again as observed experimentally \cite{mastracci-guo-2018}. \ADD{Furthermore, the normalised standard deviation 
slightly increases with temperature.} 
We conclude that the PDFs of the normal fluid in counterflow turbulence differ strikingly 
from the PDFs of velocity fluctuations in classical turbulence, which 
display a sub-Gaussian behaviour \cite{mordant-etal-2004}.  

We argue that the low-velocity regions - the 
regions which we have identified in Fig.~\ref{fig1} - are wakes
generated by the mutual friction force exerted by the vortex lines
on the normal fluid. In fact, we find that the average speed of the vortex 
lines is about one tenth of $\bar{v}_n^z$ for both
$v_{ns}^{(1)}$ and $v_{ns}^{(2)}$.
Essentially, vortices are like obstacles which 
\ADD{modify the normal fluid velocity \cite{hall-vinen-1956b,schwarz-1978}},
slowing down the normal fluid 
in the downstream region.  As $v_{ns}$ increases, the impact of these wakes
on the PDFs is less pronounced (the PDFs tend to have a more Gaussian 
character) because wakes overlap, randomizing the flow.

\begin{figure*}[htbp]
  \centering
  \includegraphics[width=.31\textwidth]{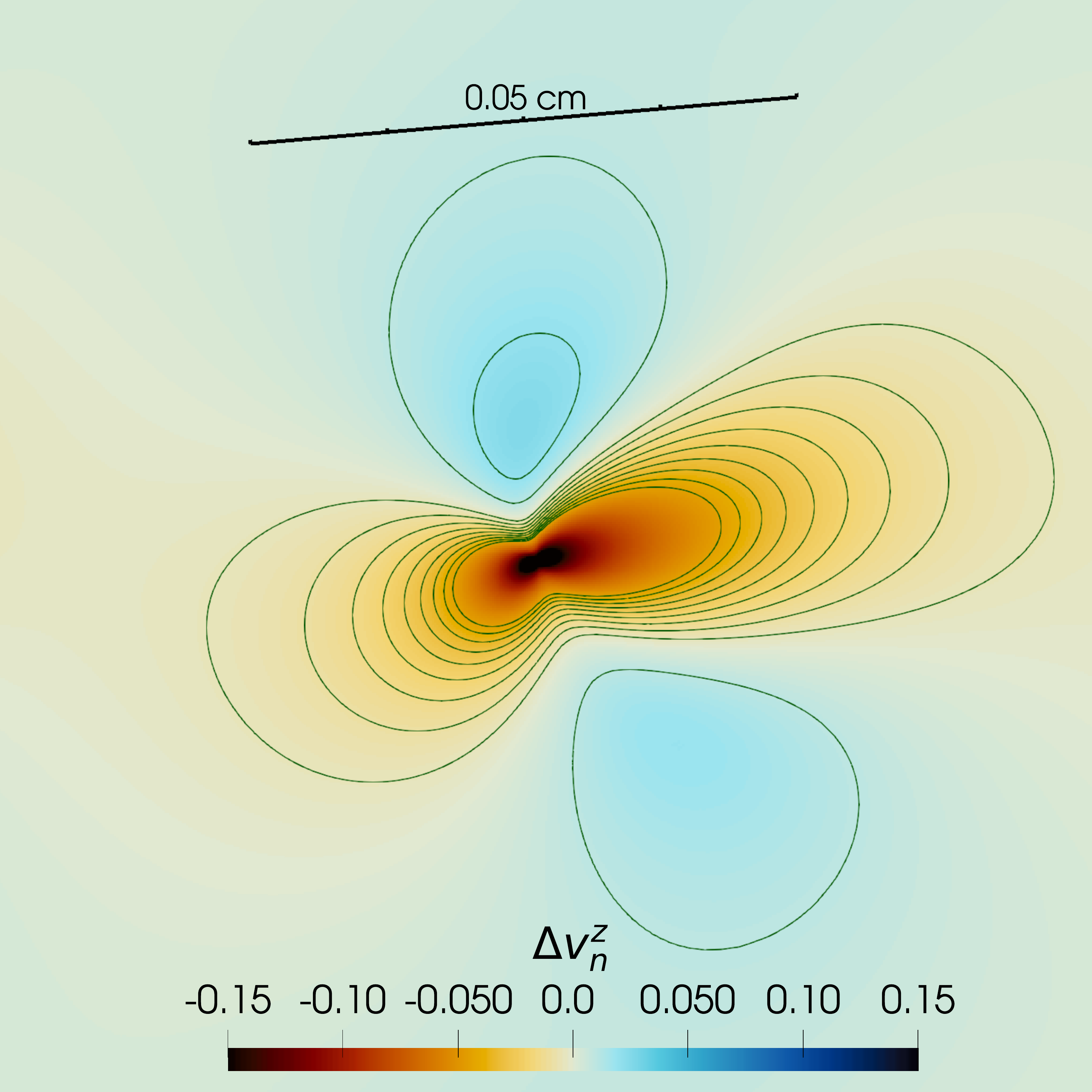}\hspace{0.02\textwidth}
  \includegraphics[width=.31\textwidth]{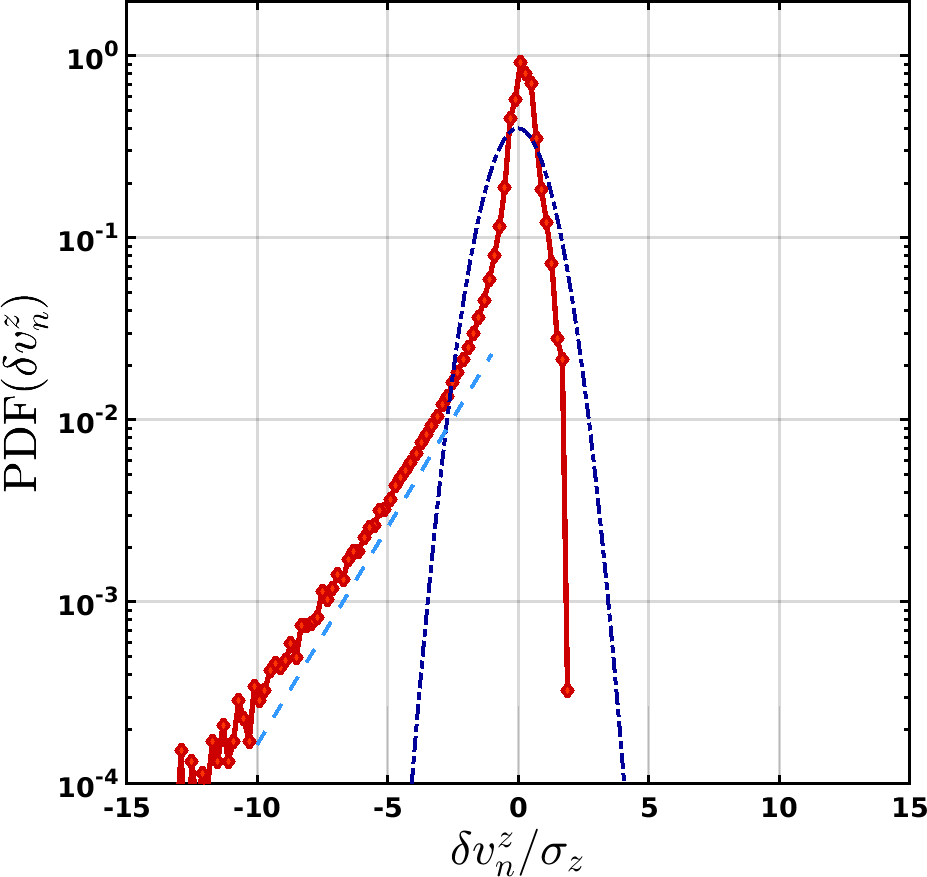}\hspace{0.02\textwidth}
  \includegraphics[width=.31\textwidth]{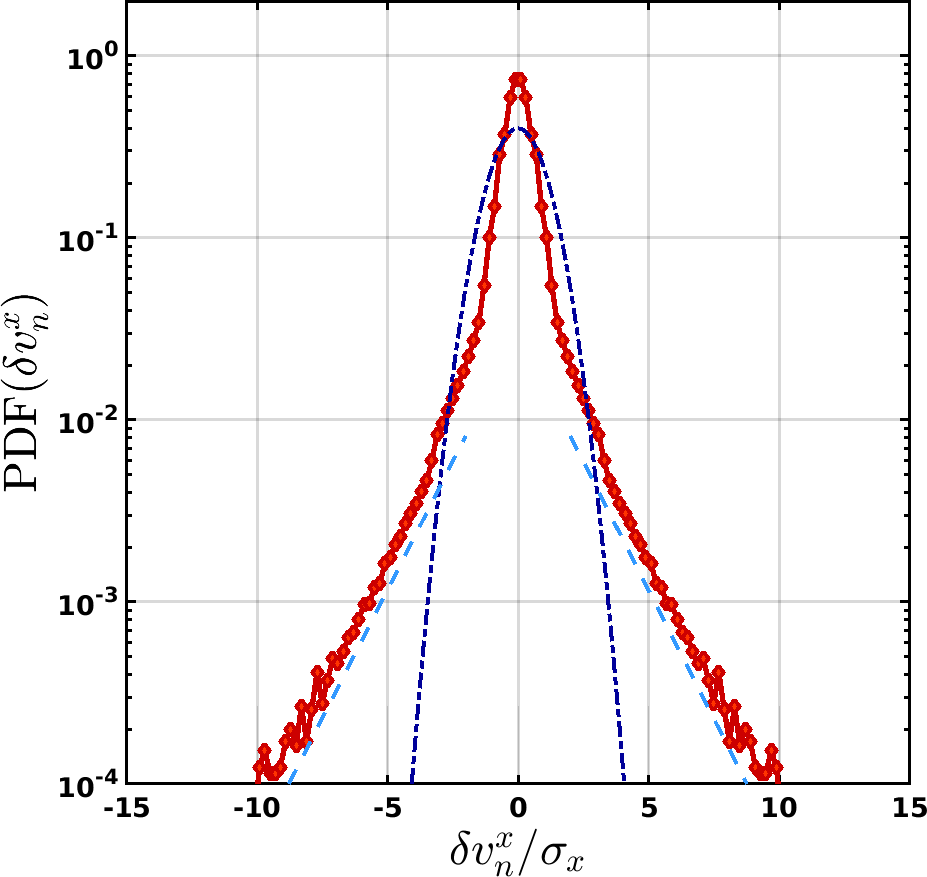}
\caption{
{\bf Single vortex wake.}
Temperature $T=1.5~\rm K$ and $v_{ns}^{(1)}=0.27 \rm cm/s$
    Left column: relative normal fluid streamwise velocity 
    fluctuations $\Delta v_n^z=\delta v_n^z/\bar{v}_n^z=(v_n^z(x,z)-\bar{v}_n^z)/\bar{v}_n^z$ 
    vs $x$ and $z$ at fixed $y_0$.  
    The ruler indicates the direction of the relative velocity of the single vortex line with respect to the normal fluid. 
    Centre column: PDF$(\delta v_n^z)$ vs $\delta v_n^z/\sigma_z$, where $\delta v_n^z=v_n^z-\bar{v}_n^z$ and $\sigma_z$
    is the standard deviation. 
    Right column: PDF$(\delta v_n^x)$ vs $\delta v_n^x/\sigma_x$, where $\delta v_n^x=v_n^x$ ($\bar{v}_n^x=0$) and $\sigma_x$
    is the standard deviation. 
    Gaussian distributions are showed in dot-dashed dark blue line for reference.
    The dashed cyan lines are exponential functions to guide the eye.
}
\label{fig3}
\end{figure*}

In order to support our argument, we perform a numerical experiment in a simpler set-up, which clarifies the physics: 
a single, straight superfluid vortex oriented in the positive $y$-direction 
in the presence of counterflow in the $z$ direction, at the same 
values of $T$ and $v_{ns}$ used for Fig.~\ref{fig1}. We find that,
after a short transient, the vortex moves at constant relative velocity 
with respect to the normal fluid.  In a turbulent tangle,
vortices move at a relatively constant velocity with respect 
to the normal fluid only for a fraction of the time between two successive reconnections $\tau_r$, which can be estimated using the vortex line density $L$ as $\tau_r\, \approx \, 6\pi/|\kappa L \ln (L^{1/2} a_0)|$ \cite{barenghi-samuels-2004}. Consequently, we consider the pattern of the normal fluid 
past the single vortex at time $t \approx \tau_r/10$. Fig.~\ref{fig3} displays the results of this numerical experiment using the counterflow velocity $v_{ns}^{(1)}$ (equivalent figures for $v_{ns}^{(2)}$ are provided in the Supplemental Information \cite{SM}).

The left panel of Fig.~\ref{fig3}, 
where contours lines are drawn at constant values of {$(v_n^z(x,z)-\bar{v}_n^z)/\bar{v}_n^z$,}
shows that the normal fluid wake is almost parabolic.
The size of the wake, computed as a velocity weighed distance from the vortex \cite{SM}, is 
$w^{(1)} = 190 \mu \rm m$ for $v_{ns}=v_{ns}^{(1)}$ and $w^{(2)} = 140 \mu \rm m$ for $v_{ns}=v_{ns}^{(2)}$, almost two orders of magnitude larger than the normal fluid vorticity dipoles
\cite{galantucci-krstulovic-etal-2023}.
The misalignment of the direction of the relative velocity between the vortex and the normal fluid (indicated by the ruler in Fig.~\ref{fig3} (left)) 
and the axis of the wake arises from the Iordanskii force (see Methods).
In the central and right panels of Fig.~\ref{fig3}, we report the PDFs of the streamwise and spanwise normal fluid velocity fluctuations in our single vortex numerical experiment. 
We observe that the PDFs (centre and right panels) mimic the behaviour of the corresponding PDFs computed for the turbulent counterflow at 
the same counterflow velocity $v_{ns}^{(1)}=0.27 \rm cm/s$ reported in Fig.~\ref{fig2} 
(red curves): skewed and/or with wide exponential tails. 

The dipole and the parabolic wake induced by 
a vortex onto the normal fluid recall the classical
Oseen solution of the slow viscous flow past a cylinder of radius $a_0$ 
equations \cite{lamb-1932}. Indeed, the assumptions behind the Oseen solution
are well satisfied: the vortex is locally straight (its radius of curvature is $R_c \, \approx \, \ell \, \gg a_0$) hence the flow is two-dimensional, the Reynolds number $Re=\bar{v}_n^z a_0/\nu_n$ (where $\nu_n$ is the kinematic viscosity) is very small: $Re^{(1)}=0.3\times 10^{-5}$ and $Re^{(2)}=1\times 10^{-5}$, and the scales at which we probe the flow are much larger than the cylinder's
radius. Focusing on the far field solution at distances larger than the Oseen scale $a_0/Re$, we recover both the vortex dipole and the parabolic wake in intensity and shape, and the skewed/exponential velocity fluctuation PDFs \cite{SM}.
\ADD{The far wake Oseen solution also provides a very good estimate for the velocity fluctuations reported in Fig.~\ref{fig2}.}

Our results thus suggest that the flow of the normal fluid in a turbulent helium counterflow can be described as the superposition of a
uniform background flow generated  by the heater and flow disturbances 
(consisting of small vortex dipoles and large almost parabolic wakes) 
generated by the vortices via the friction force.
The less pronounced tails of the normal fluid velocity
fluctuations for $v_{ns}^{(2)}=0.94 \rm cm/s$ (Fig.~\ref{fig2}, green curves) 
arise from the fact that the wake $w^{(2)}$ is larger than the average inter-vortex spacing 
$\ell^{(2)}=94 \mu \rm m$,  hence wakes overlap randomising the flow. On the contrary,
for $v_{ns}^{(1)}=0.27 \rm cm/s$,  $w^{(1)} < \ell^{(1)}=370 \mu \rm m$, 
the wakes tend to be separated, and the velocity statistics of the normal fluid
(Fig.~\ref{fig2}, red curves) echo the case of an isolated vortex (Fig.~\ref{fig3}, 
central and right panels). 

\begin{figure}[htbp]
  \centering
  \includegraphics[width=.4\textwidth]{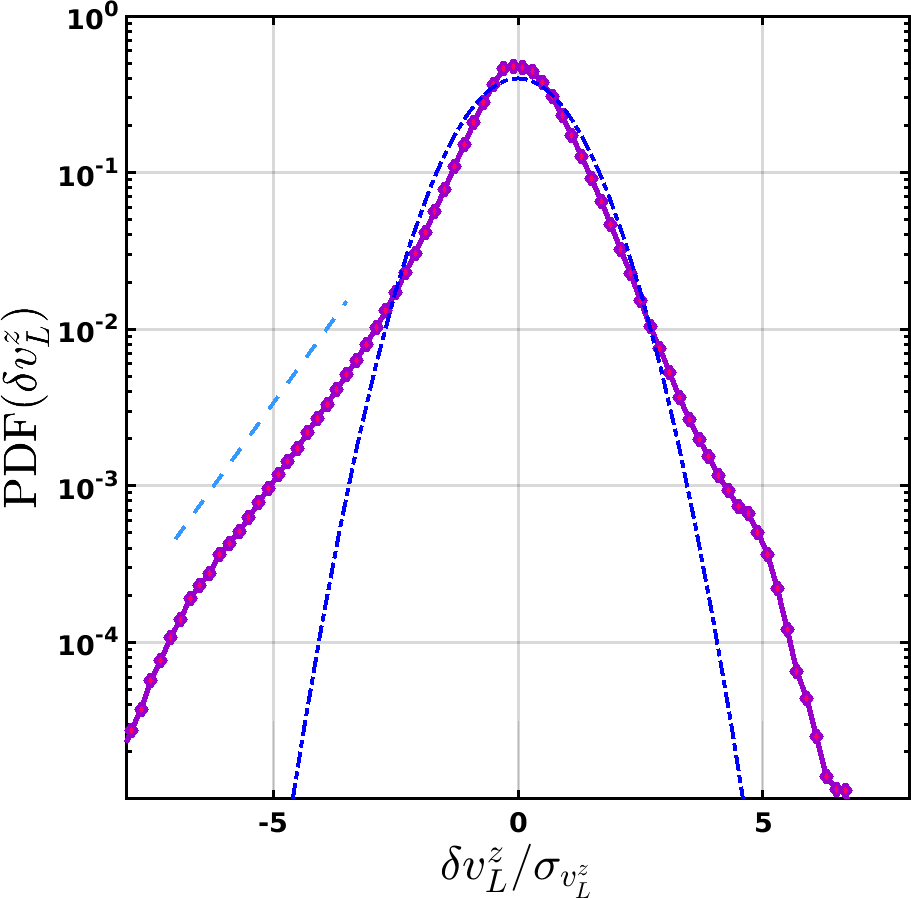}
    \caption{
     {\bf Vortex statistics.} 
     Parameters: $T=1.5~\rm K$, $v_{ns}=v_{ns}^{(2)}=0.94~\rm cm/s.$
     Distribution of streamwise vortex velocity fluctuations
     ${\rm PDF}(\delta v_L^z)$ vs $\delta v_L^z/\sigma_{v_L^z}$ where
     $\sigma_{v_L^z}$ is the standard deviation of $v_L^z$. 
     The dot-dashed blue line is the Gaussian fit, 
     and the cyan dahed line is a guide to the eyes to highlight the tail.
    }
\label{fig4}
\end{figure}  

The natural question is whether the normal fluid wakes that we predict
can be detected in experiments. 
Non-distructive visualisation of the turbulent normal fluid has been achieved
using metastable helium molecules \cite{Guo-etal-PRL-2010}, but has not yet
probed the length scales considered here.
However, measurements obtained via 
the PTV visualization technique may show the signature of the wakes. 
Indeed, a recent PTV experiment at large counterflow velocities reported a
left-skewed distribution for the streamwise
velocity of likely trapped particles (the so-called ``slow'' particles in Fig.~10 of 
Ref.~\cite{svancara-etal-2021}), similar to the normal fluid streamwise velocity
distributions that we calculate (our Fig.~\ref{fig2} (left)).

A similar left-skewed distribution (see Fig.~\ref{fig4}) is recovered when
calculating at $v_{ns}=v_{ns}^{(2)}$ (as experiments are performed at large $v_{ns}$) 
the distribution of the streamwise vortex velocity fluctuations
$\delta v_L^z=v_L^z - \bar{v}_L^z$, where $\mathbf{v}_L$ is the vortex velocity, confirming that the motion of particles trapped on vortices may 
be indeed approximated with the motion of the vortices themselves 
\cite{mastracci-guo-2018,yui-etal-2022}.
It is possible to show \cite{SM} that the skewness which characterises
$\delta v_n^z$ (Fig.~\ref{fig2} (left)) contributes to the enhance the skewness of $\delta v_L^z$ (Fig.~\ref{fig4}): 
the experimental signature of the normal fluid wakes is the left-skewed distribution of the streamwise velocity of
likely trapped particles in PTV experiments \cite{svancara-etal-2021}.

Finally, it is interesting to notice that the original one-way coupled
theory of Schwarz \cite{schwarz-1988} (where 
$\vvec_n= \bar{\vvec}_n$, prescribed \textit{a priori})
predicts that, at $T=1.5~\rm K$,
the vortex tangle drifts, with respect to the superfluid applied flow, at
average velocity $\bar{v}_L^z -\bar{v}_s^z  \approx 0.35 {v}_{ns}$ in the
direction of $v_n^z$ \cite{schwarz-1978}, as a result of the drag exerted by the normal fluid. 
A subsequent experiment \cite{awschalom-etal-1984} found that the average drift velocity of the vortices is smaller, reporting the upper limit $\bar{v}_L^z -\bar{v}_s^z  <0.2 {v}_{ns}$ for this temperature, \ADD{which has been confirmed by more recent numerical and theoretical investigations of the Schwarz model 
\cite{Kondaurova-Lvov-Procaccia-2014,nemirovskii-2013}}. 
Our two-way coupled simulation finds $\bar{v}_L^z -\bar{v}_s^z\le 0.15 {v}_{ns}$,
in good agreement with the experiment \cite{awschalom-etal-1984}, showing that vortices see a slower 
normal fluid as a result of the wakes. \ADD{The issue of the drift of the vortex tangle remains nevertheless open, 
as it entangles different contributions from anisotropy, two-fluid coupling and wakes.}

In conclusion, our numerical experiments predict that counterflow
turbulence can be described as the superposition of (i) the superflow
and the normal flow (here uniform) which are imposed
by the heater, (ii) superfluid velocity fluctuations created by
the turbulent vortex lines, and (iii)
local perturbations of the normal fluid velocity induced by the 
friction. 
Two decades ago theory predicted \cite{kivotides-barenghi-samuels-2000}
that the vortex lines
are surrounded by small dipolar vorticity structures in the normal fluid.
The very small scale of these structures (comparable to the
scale of the particles used for PTV visualization) has so far
prevented any direct observation. The novel feature revealed by the work
presented here is that, alongside the dipolar structures, the vortex lines also
induce normal fluid wakes  which can be larger than
the average intervortex distance and affect the velocity
statistics of the normal fluid, as the experiments that we have discussed
suggest. \ADD{One could therefore inquire if such large-scale wakes could also affect thermodynamic quantities  \cite{nemirovskii2024dependence}, leading to interesting new phenomena.}







%

\subsection*{Methods}
\textbf{The FOUCAULT model.}
Our model considers the the superfluid as a collection ${\cal L}$ of vortex lines that advect each other following Biot-Savart integrals. 
Vortex lines are parametrised employing arclength $\xi$ and time $t$, \textit{i.e.} the position of the vortex filaments is given by $\bolds(\xi,t)$.
Regularisation of the integrals and reconnections are built following the Schwarz's vortex filament model \cite{schwarz-1988}. The vortex lines interact with the normal fluid through mutual friction coefficients $\beta$ and $\beta'$ as follows
\begin{equation}
  \mathbf{v}_L = \dot{\bolds}(\xi,t)=\frac{\partial \bolds}{\partial t}=
  \boldv_{\rm s_\perp}+\beta \bolds' \times \boldv_{\rm ns}
  -\beta' \bolds' \times (\bolds' \times \boldv_{\rm ns}),
  \label{eq:VF}
\end{equation}
where $\bolds'=\partial\bolds/\partial \xi$ is the unit tangent vector, the subscript `$\perp$' indicates the component of the superfluid velocity 
lying on a plane orthogonal to $\mathbf{s}'$ and $\boldv_{\rm ns}=\boldv_{\rm n}-\boldv_{\rm s}$ at $\bolds$, with $\boldv_{\rm s}$ the superfluid velocity
given by the Biot-Savart integral
\begin{equation}
  \boldv_{\rm s}({\bf s},t)=\overline{  \boldv}_{\rm s}+\frac{\kappa}{4 \pi} \oint_{\cal L}
  \frac{\bolds'(\xi,t) \times ({\bf s}-\bolds(\xi,t))}
  {\vert {\bf s} -\bolds(\xi,t)\vert^3 }d\xi .
  \label{eq:BS}
\end{equation}
In turn, the normal fluid satisfies the incompressible Navier-Stokes equations
\begin{eqnarray}
\displaystyle
\frac{\partial \boldv_{\rm n}}{\partial t} + \left ( \boldv_{\rm n}\cdot \nabla \right )\boldv_{\rm n}   = 
-\frac{1}{\rho} \nabla p_{\rm n} &+& \nu_{\rm n} \nabla^2 \boldv_{\rm n} +\frac{\mathbf{F}_{\rm ns}}{\rho_{\rm n}}
\label{eq:NS} \\
\mathbf{F}_{\rm ns}=\oint_\mathcal{L}\mathbf{f}_{\rm ns}(\mathbf{s})\delta({\bf x}-\mathbf{s})\mathrm{d}\xi\,\,, &  &  \nabla \cdot \boldv_{\rm n}=0 \label{eq:incompr} \; .
\end{eqnarray}
where $\mathbf{F}_{\rm ns}$ is mutual friction force per unit volume responsible for the retroaction of quantum vortices on $\boldv_{\rm n}$, 
and
\begin{equation}
\displaystyle 
\mathbf{f}_{\rm ns} (\bolds)= - D \,\mathbf{s}' \times \left [ \mathbf{s}' \times \left ( \dot{\mathbf{s}} - \mathbf{v}_n(\bolds) \right ) \right ]
- \rho_n\kappa\; \mathbf{s}' \times \left (\dot{\mathbf{s}} - \mathbf{v}_n(\bolds)\right )
\label{eq:f_ns}
\end{equation}
its local density, where $D$ is a friction coefficient \cite{galantucci-krstulovic-etal-2020}. The second term in equation~\ref{eq:f_ns} is the Iordanskii
force: this term is responsible for the misalignment of the direction of the relative velocity between the vortex and the normal fluid 
(indicated by the ruler in Fig.~\ref{fig3} (left)) and the axis of the wake.
The total density of liquid helium is $\rho= \rho_{\rm n} + \rho_{\rm s}$, with $\rho_{\rm n}$ and $\rho_{\rm s}$ are respectively the normal fluid and superfluid densities. Finally, $p_{\rm n}$ is the effective pressure, 
and $\nu_{\rm n}$ the kinematic viscosity of the normal fluid. The mean normal fluid velocity $\overline{  \boldv}_{\rm n}$ is directly imposed during the evolution.

A complete description of the Fully cOUpled loCAl model of sUperfLuid Turbulence (FOUCAULT), including its numerical implementation, can be found in \cite{galantucci-krstulovic-etal-2020}.

\textbf{Numerical details.} The Navier-Stokes equations are solved in a periodic domain using a standard pseudo-spectral code. In all simulations we use a grid of $512^3$ grid points. The vorte filament model is evolved using a Tree-Algortihm method thats allows for an important speed up of the Biot-Savart computations. The number of discretisation points used to represent the vortex lines is adapted during the simulations, and fluctuates in the steady state around $700$ and $10^4$ for $v_{ns}^{(1)}$ and $v_{ns}^{(2)}$, respectively.

\subsection*{Data availability}
Data is availability upon request.

\subsection*{Acknowledgements}
We acknowledge discussions with Profs Marco La Mantia, Mathieu Gibert and Wei Guo. G.K. acknowledges support of Agence Nationale de la Recherche through the project QuantumVIW ANR-23-CE30-0024-02.

\subsection*{Author contributions} 
L.G ran the simulations and processed the data. All authors contributed to writing the paper.

\subsection*{Competing interests}
The authors declare no competing interests.



\newpage
\newpage

\subsection*{Supplementary Information}
\textbf{Single vortex wake at $v_{ns}=v_{ns}^{(2)}=0.94 \rm cm/s$}\\[1mm]

\begin{figure*}[htbp]
  \centering
   \includegraphics[width=.31\textwidth]{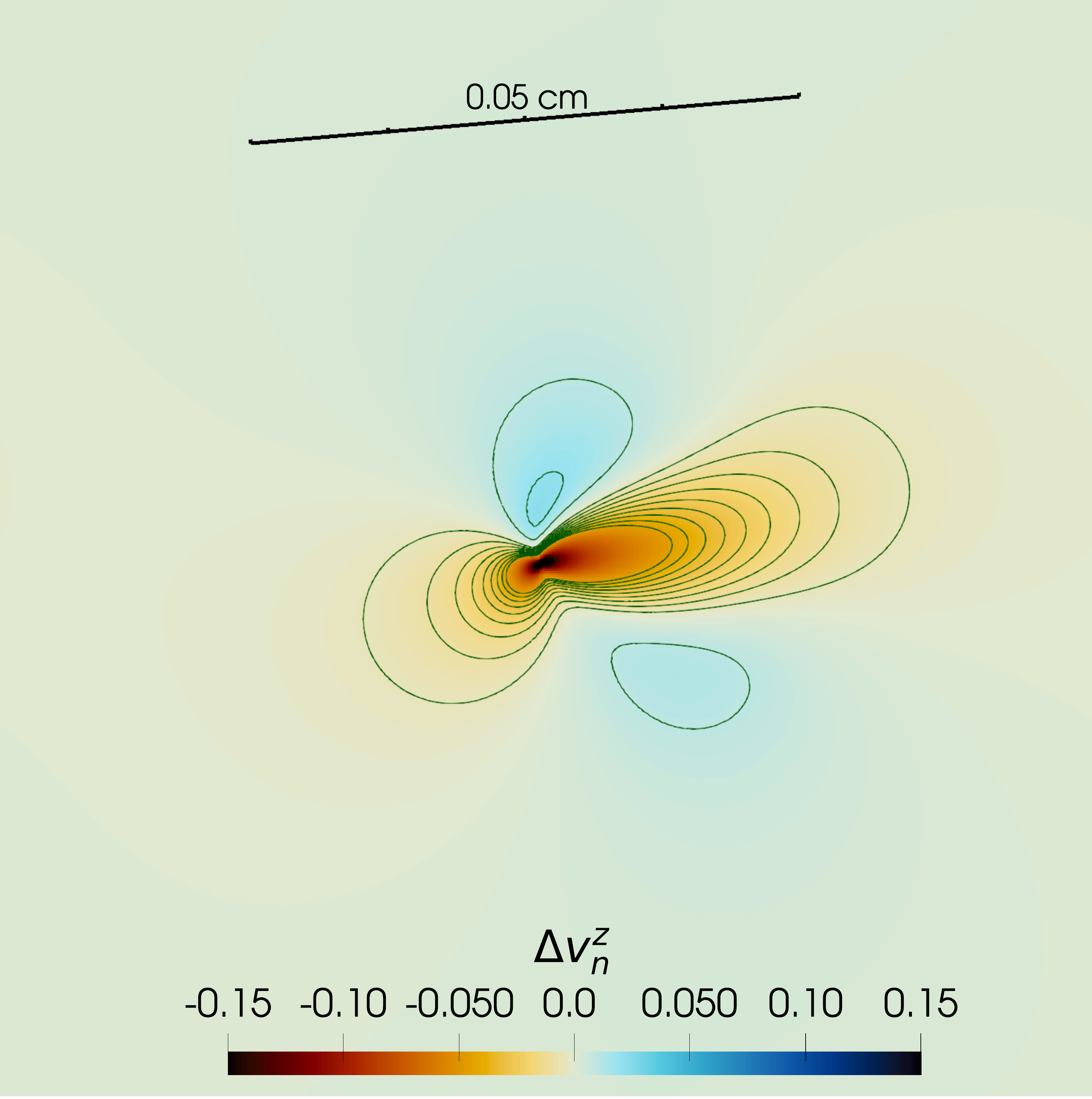}\hspace{0.02\textwidth}
   \includegraphics[width=.31\textwidth]{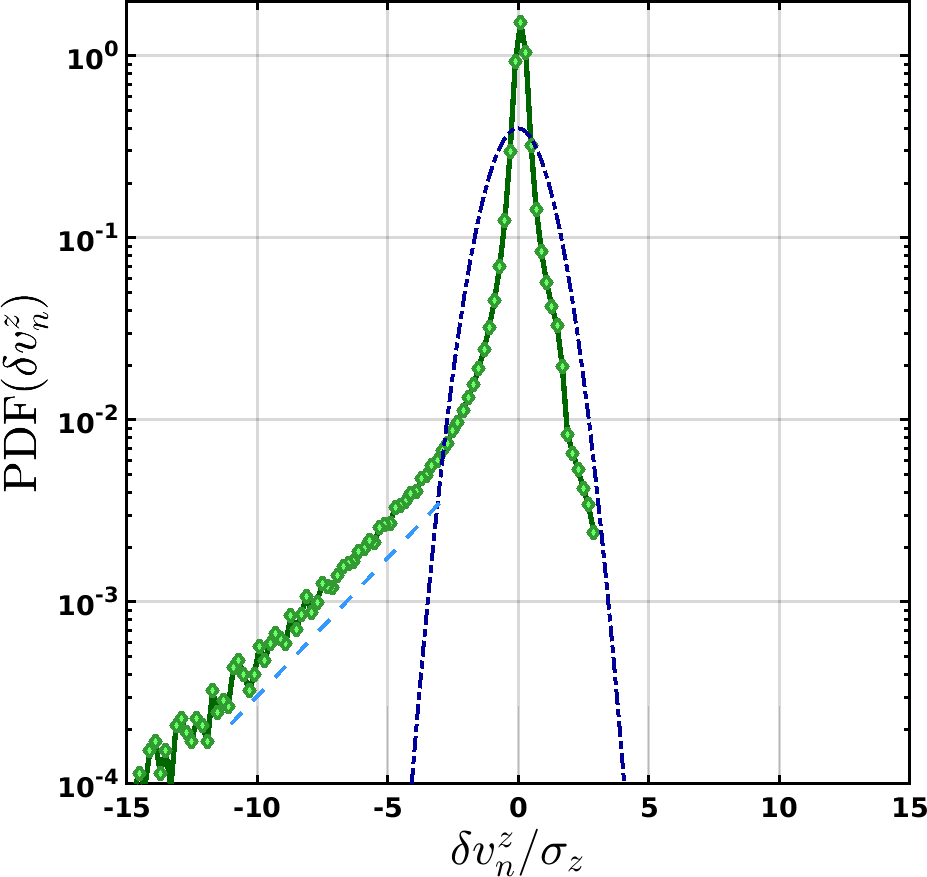}\hspace{0.02\textwidth}
   \includegraphics[width=.31\textwidth]{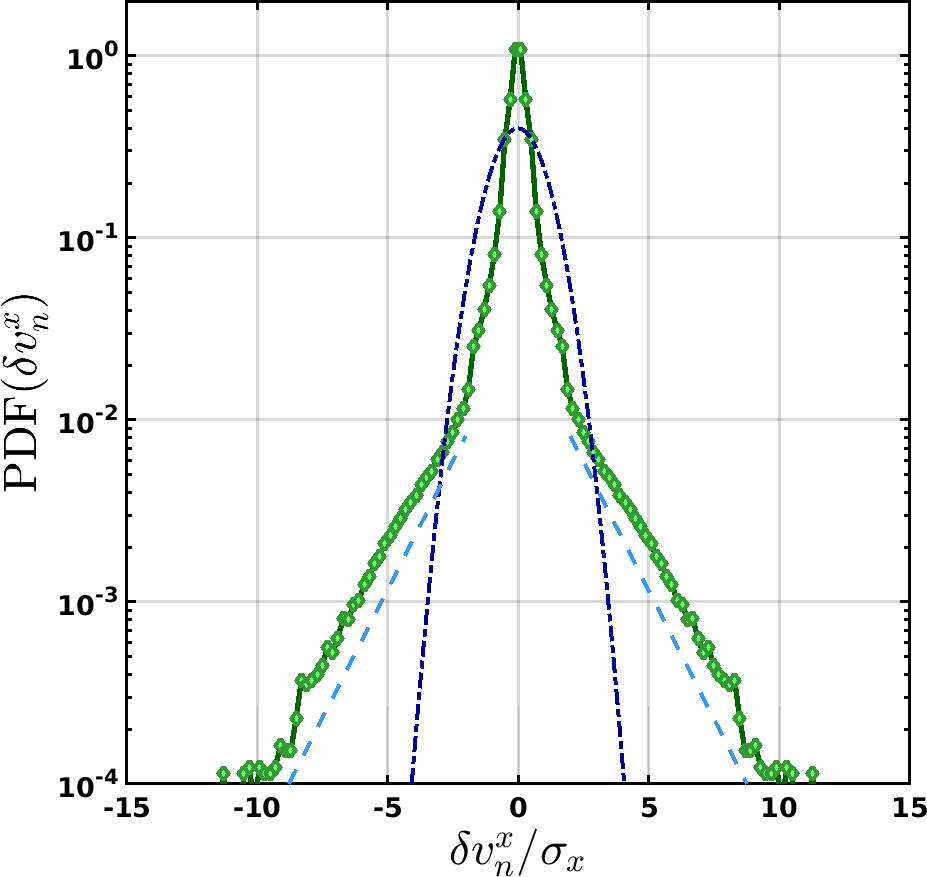}
\caption{
{\bf Single vortex wake.}
Temperature $T=1.5~\rm K$ and $v_{ns}^{(2)}=0.94 \rm cm/s$
    Left column: relative normal fluid streamwise velocity 
    fluctuations $\Delta v_n^z=\delta v_n^z/\bar{v}_n^z=(v_n^z(x,z)-\bar{v}_n^z)/\bar{v}_n^z$ 
    vs $x$ and $z$ at fixed $y_0$.  
    The ruler indicates the direction of the relative velocity of the single vortex line with respect to the normal fluid. 
    Centre column: PDF$(\delta v_n^z)$ vs $\delta v_n^z/\sigma_z$, where $\delta v_n^z=v_n^z-\bar{v}_n^z$ and $\sigma_z$
    is the standard deviation. 
    Right column: PDF$(\delta v_n^x)$ vs $\delta v_n^x/\sigma_x$, where $\delta v_n^x=v_n^x$ ($\bar{v}_n^x=0$) and $\sigma_x$
    is the standard deviation. 
    Gaussian distributions are showed in dot-dashed dark blue line for reference.
    The dashed cyan lines are exponential functions to guide the eye.
}
\label{figS1}
\end{figure*}

The pattern of the normal fluid past a single vortex at counterflow velocity
$v_{ns}=v_{ns}^{(1)}=0.27~\rm cm/s$ was shown in Fig.~\ref{fig3} of the main text. For completeness,
Fig.~\ref{figS1} displays the wake generated by a single, straight superfluid vortex oriented 
in the positive $y$-direction in the presence of counterflow in the $z$-direction at counterflow 
velocity $v_{ns}=v_{ns}^{(2)}=0.94 \rm cm/s$ at $t\approx \tau_r/10$.
In the left panel, we report the relative normal fluid streamwise velocity fluctuations 
$\Delta v_n^z=\delta v_n^z/\bar{v}_n^z=(v_n^z(x,z)-\bar{v}_n^z)/\bar{v}_n^z$ vs $x$ and $z$ at fixed $y_0$. 
In the centre (right) panel we report the PDFs of the streamwise (spanwise) normal fluid velocity fluctuations $\delta v_n^z$ ($\delta v_n^x$). 
With respect to the spanwise ($x$) fluctuations (Fig.~\ref{figS1}, right), the PDF has been computed taking also into account the 
fluctuations generated by a vortex oriented in the negative $y$-direction (this is not necessary for the streamwise component as in this circumstance
the fluctuations generated coincide).
  
\vspace{2cm}

\textbf{Calculation of wake size in the single vortex simulation}\\[1mm]

For both counterflow velocities $v_{ns}^{(1)}$ and $v_{ns}^{(2)}$ we compute the size of the wake at $t\approx \tau_r/10$, employing the two-dimensional 
normal fluid velocity field resulting from the motion of a single, straight superfluid vortex in presence of a counterflow. The 
corresponding relative streamwise velocity fluctuations {$(v_n^z(x,z)-\bar{v}_n^z)/\bar{v}_n^z$} have been reported in Fig.~\ref{fig3} (left) in the main \red{text} 
and Fig.~\ref{figS1} (left) in the Supplementary Information for $v_{ns}^{(1)}$ and $v_{ns}^{(2)}$, 
respectively. The size of the wake, $w$, is \red{defined} as follows:

\begin{equation}
\displaystyle 
w=\sqrt{\frac{\displaystyle \int_\mathcal{S}\! |\mathbf{x}-\bolds_0|^2 |\delta\tilde{{\bf v}}_n|^2 dxdz}{\displaystyle\int_\mathcal{S}\!\! |\delta\tilde{{\bf v}}_n|^2  dxdz}} \;\; ,
\label{eq:wake_size}
\end{equation}
where $\mathcal{S}$ is the $x$-$z$ two-dimensional domain orthogonal to the superfluid vortex 
at fixed $y_0$ (\text{i.e.} the domain reported in Fig.~\ref{fig3} (left) in the main \red{text}
and Fig.~\ref{figS1} (left) in the Supplementary Information); $\mathbf{x}=(x,z)$ 
is the general position vector; $\bolds_0$ is the intersection of the vortex 
with $\mathcal{S}$ at $t\approx \tau_r/10$; 
$\delta\tilde{{\bf v}}_n = (v_n^x,\red{v_n^y},v_n^z -\bar{v}_n^z)$.

\vspace{2cm}

\textbf{Oseen solution}\\[1mm]

\begin{figure*}[htbp]
  \centering
  \includegraphics[width=.45\textwidth]{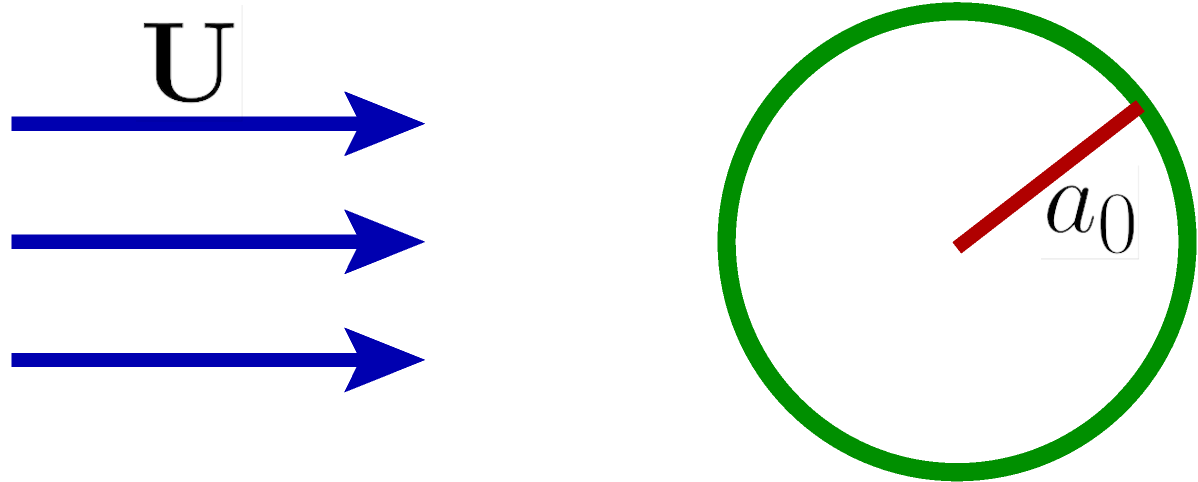}
    \caption{
     {\bf Schematic rendering of the two-dimensional flow past a circle}.
    }
\label{fig:Oseen}
\end{figure*}

In two-dimensions, the Stokes equation describing the steady flow of a viscous fluid past a circle 
at low Reynolds number has no solution: it is impossible to
enforce simultaneously no-slip boundary condition on the circle \red{the condition of}
and uniform flow at infinity (Stokes'paradox). In order to determine the flow
of the fluid, it is \red{necessary} to employ Oseen's equation \cite{lamb-1932} including a 
linearised convective term. With reference to Fig.~\ref{fig:Oseen},
the velocity of the viscous fluid $\mathbf{V}$ can be expressed in terms of the velocity 
of the fluid at infinity $\mathbf{U}$ and a perturbation $\mathbf{v}$, 
as follows $\mathbf{V}=\mathbf{U} + \mathbf{v}$, where $\mathbf{U}=U\hat{\mathbf{k}}$, 
$\hat{\mathbf{k}}$ being the unit vector in the $z$ (horizontal) direction
($x$ is the vertical direction). At large distances $r \gg a_0$, the distances we are 
interested in, $\mathbf{V} \approx \mathbf{U}$
(or, equivalently, $v \ll U$) and hence the convective term may be approximated as follows, 
$ \displaystyle(\mathbf{V}\cdot\nabla)\mathbf{V} \approx U\frac{\partial}{\partial z}\mathbf{v}$ 
leading to the Oseen equation for the perturbation $\mathbf{v}$:

\begin{equation}
\displaystyle 
U\frac{\partial}{\partial z}\mathbf{v} = -\frac{1}{\rho}\nabla p + \nu \nabla^2 \mathbf{v}.
\label{eq:oseen}
\end{equation}

\begin{figure*}[htbp]
  \centering
  \includegraphics[width=.6\textwidth]{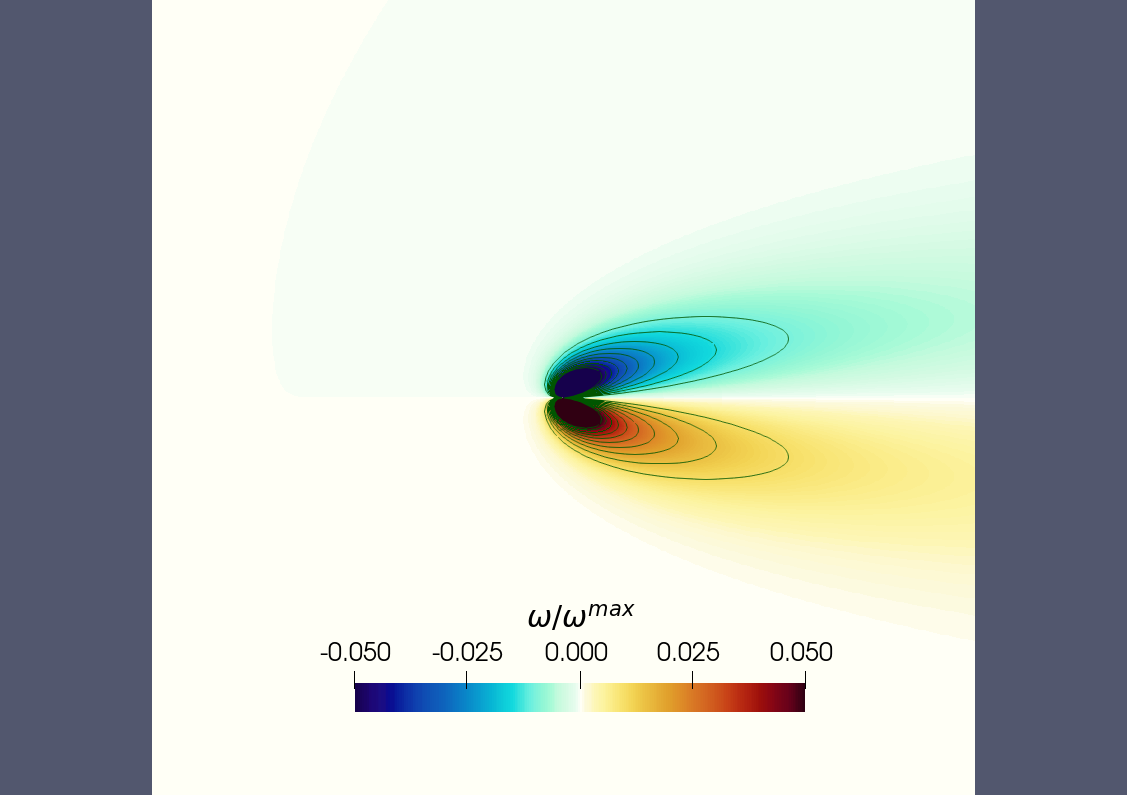}
    \caption{
     {\bf Far field Oseen solution:} vorticity field normalised by its maximum value $\omega/\omega_{max}$. The size of the domain is 0.1 cm.
    }
\label{fig:oseen_far_field_w}
\end{figure*}

The non-dimensional far field solution of equation (\ref{eq:oseen}), \textit{i.e.} the solution for $r \gg a_0/Re$, is in polar coordinates \cite{lamb-1932}

\begin{eqnarray}
\displaystyle
v'_r & = & \frac{C_0}{2}\sqrt{\frac{\pi}{\tilde{r}}} \left (1+ \cos\theta \right )e^{ -\frac{\tilde{r}}{2}(1-\cos\theta)} - \frac{C_0}{\tilde{r}}, \nonumber \\[2mm]
v'_\theta & = & -\frac{C_0}{2}\sqrt{\frac{\pi}{\tilde{r}}} \sin\theta\, e^{ -\frac{\tilde{r}}{2}(1-\cos\theta)} 
\label{eq:oseen_far_field_v} \; ,
\end{eqnarray}

\noindent
where $\displaystyle\tilde{r}=r\frac{Re}{a_0}$ is the Oseen \red{non-dimensional}
variable (the far field corresponds to $\tilde{r}\gg 1$), 
\begin{equation}
C_0=-\frac{2}{0.5-\gamma -\log (\epsilon/2)},
\end{equation}
\noindent
$\gamma=0.5772$ is the Euler-Mascheroni constant and $v'=v/U$. 
The resulting \red{far-field} vorticity $\omega$ is given by the following expression

\begin{equation}
\displaystyle 
\omega = \frac{C_0 Re}{2}\sin\theta e^{ -\frac{\tilde{r}}{2}(1-\cos\theta)}\sqrt{\frac{\pi}{\tilde{r}}} \; \; ,
\label{eq:oseen_far_field_w}
\end{equation}

\noindent
whose spatial dependence is reported in Fig.~\ref{fig:oseen_far_field_w}. It is important to note 
the upstream/downstream asymmetry.

\begin{figure*}[htbp]
  \centering
  \includegraphics[width=.6\textwidth]{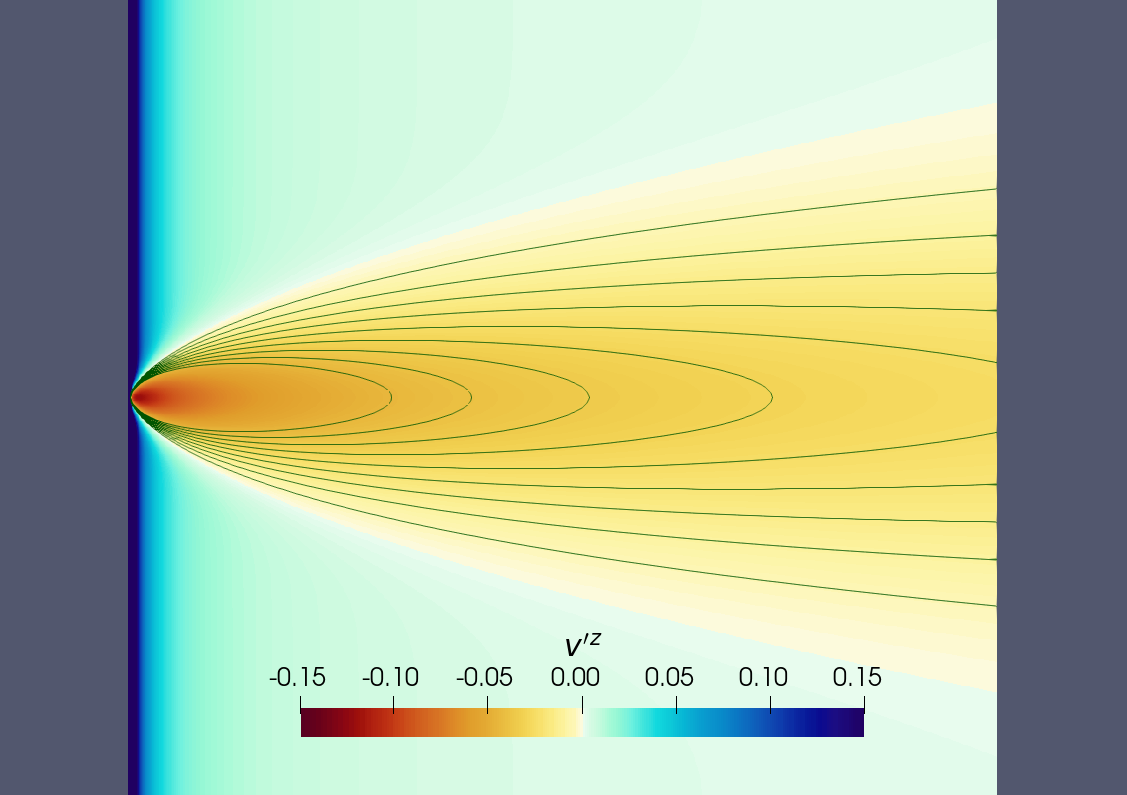}
    \caption{
     {\bf Far Wake Oseen solution:} relative magnitude of the streamwise velocity fluctuations $v'^z=v^z/U=(V^z-U)/U$. The size of the domain is 0.1 cm.
     Iso-lines are drawn at the same values as in Figs.~\ref{fig3} and \ref{figS1}.
    }
\label{fig:oseen_far_wake_v}
\end{figure*}

\begin{figure*}[htbp]
  \centering
   \includegraphics[width=.45\textwidth]{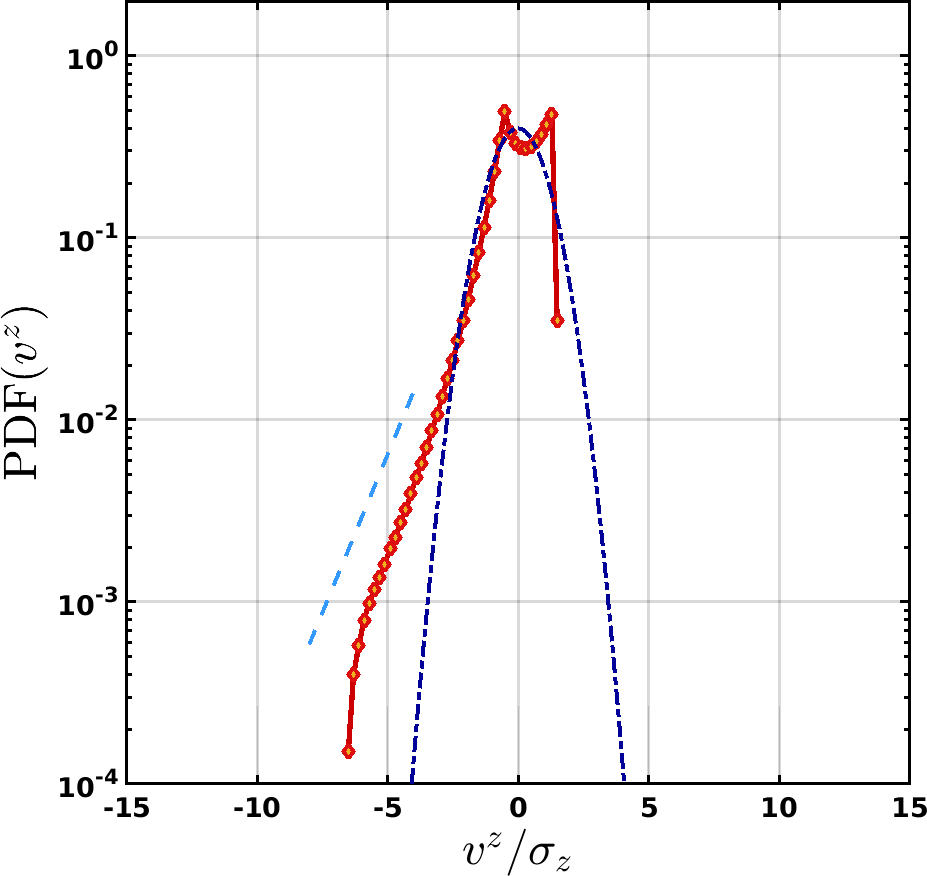}\hspace{0.02\textwidth}
   \includegraphics[width=.45\textwidth]{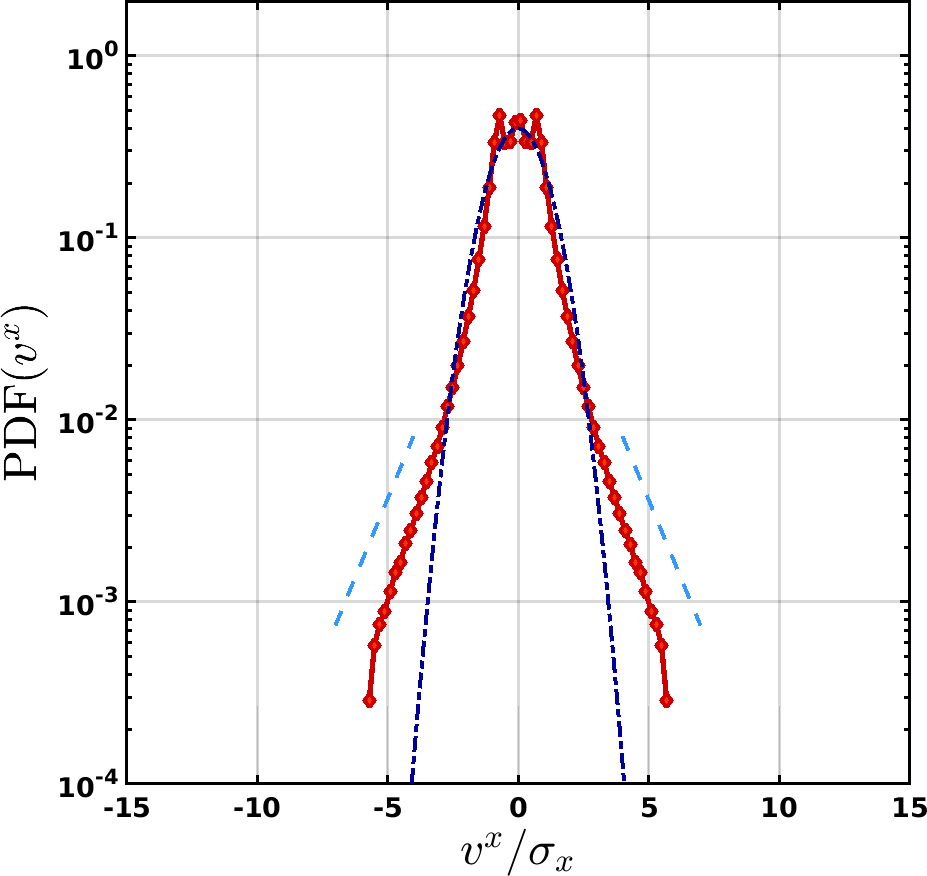}
\caption{
{\bf Far wake Oseen solution:} 
    Left: PDF$(v^z)$ vs $v^z/\sigma_z$, where $v^z$ is the streamwise velocity fluctuation of the solution of Oseen's equation in the far wake region 
    and $\sigma_z$     is the standard deviation. 
    Right: PDF$(v^x)$ vs $v^x/\sigma_x$, where $v^x$ is the spanwise velocity fluctuation of the solution of Oseen's equation in the far wake region 
    and $\sigma_x$     is the standard deviation. 
    Gaussian distributions are showed in dot-dashed dark blue line for reference.
    The dashed cyan lines are exponential functions to guide the eye.
}
\label{fig:pdf_oseen_v}
\end{figure*}

The far field solution of Oseen's equation (\ref{eq:oseen_far_field_v}) can be employed 
to determine the spatial dependence of the flow 
in the far wake, \textit{i.e.} for $\tilde{r}\gg 1$ and $\theta \ll 1$. 
In this region the flow is described by the following expressions

\begin{eqnarray}
\displaystyle
v'_r & = & C_0\sqrt{\frac{\pi}{\tilde{r}}}\, e^{ -\frac{\tilde{r\theta^2}}{4}} - \frac{C_0}{\tilde{r}}, \nonumber \\[2mm]
v'_\theta & = & -\frac{C_0}{2}\sqrt{\frac{\pi}{\tilde{r}}} \theta\, e^{ -\frac{\tilde{r\theta^2}}{4}} 
\label{eq:oseen_far_wake_v} \; ,
\end{eqnarray}
\noindent
whose streamwise component $v'^z = (V^z-U)/U$ is reported in Fig.~\ref{fig:oseen_far_wake_v}. The iso-lines in Fig.~\ref{fig:oseen_far_wake_v}
are drawn at the same values of the relative magnitude of streamwise velocity fluctuations employed 
in Fig.~\ref{fig3}. 
\red{If we compare} the linear sizes of the wakes (\textit{e.g} the linear size of the region where $v'^z < -0.05$)
we find very good agreement between
the analytical solution of Oseen's equation and our numerical results. 
Furthermore, if we compute the PDFs of the streamwise and spanwise velocity fluctuations,
$v^z$ and $v^x$, respectively reported in Fig.~\ref{fig:pdf_oseen_v} (left) and (right), 
in the far wake (for $\tilde{r} > 10$ and $\theta \ll 10^\circ$), we recover the same 
characteristics observed for the normal fluid in our counterflow turbulence simulations 
and in the analysis of the single vortex wake: 
the PDFs are skewed and/or with exponential tails.
This similarity supports our conclusion \red{that} the normal fluid in a turbulent helium counterflow \red{is}
the superposition of a
uniform background flow generated  by the heater and flow disturbances 
(consisting of small vortex dipoles and large almost parabolic wakes) 
generated by the vortices via the friction force.

\vspace{2cm}

\textbf{Skewness of vortex velocity $\mathbf{v}_L$}\\[1mm]

\red{Consider equation (\ref{eq:VF})}.  
For $T\ge 1.5$K (the temperature range where the $\rho_n/\rho > 0.1$) \red{we have}
$|\beta'| > \beta > 0$, \red{hence}
the streamwise vortex velocity $v_L^z$ can be written as follows:

\begin{eqnarray}
  v_L^z(\bolds) & = & (1-|\beta'|) (\bar{v}_s^z + v_{_{BS}}^z(\bolds)) + |\beta'| v_n^z(\bolds) \nonumber \\[2mm]
        & = & \varepsilon_1 (\bar{v}_s^z + v_{_{BS}}^z(\bolds)) + \varepsilon_2 v_n^z(\bolds)\; \; ,
  \label{eq:VF_z}
\end{eqnarray}
\noindent
where $v_{_{BS}}^z(\bolds)$ is the streamwise component of the Biot-Savart integral 
in equation (\ref{eq:BS}), $\varepsilon_1=1-|\beta'|=0.88$, 
$\varepsilon_2=|\beta'|=0.12$, 
$\varepsilon_1+\varepsilon_2=1$, $\varepsilon_1$ and $\varepsilon_2$ depend on temperature $T$ 
and Reynolds number and as temperature
increases $\varepsilon_1 \rightarrow 0$ and $\varepsilon_2 \rightarrow 1\,$. 
Values of $\varepsilon_1$ and $\varepsilon_2$ reported
are evaluated at $T=1.5$K \cite{galantucci-krstulovic-etal-2020}, and \red{the velocities}
$v_{_{BS}}^z$ and $v_n^z$ are evaluated on the vortex positions $\bolds$. 

Equation~(\ref{eq:VF_z}) implies that, \red{in the first approximation},
the streamwise vortex velocity fluctuation $\delta v_L^z$ has the following
expression:

\begin{equation}
\displaystyle 
\delta v_L^z (\bolds)= \varepsilon_1 \delta v_{_{BS}}^z(\bolds) + \varepsilon_2 \delta v_n^z(\bolds)
\label{eq:VF_z_fluct}
\end{equation}

To quantify the asymmetry of the PDF of $\delta v_L^z$ (reported in Fig.~\ref{fig4} and echoing the experimental non symmetric distribution of the streamwise velocity of particles likely trapped on vortices \cite{svancara-etal-2021}), the relevant quantity to compute is the skewness of $v_L^z$ defined as 
$\mathcal{S}(v_L^z) = \overline{\left [ \delta v_L^z (\bolds) \right ]^3}$ which, following equation~(\ref{eq:VF_z_fluct}), depends on 
$\mathcal{S}(v_{_{BS}}^z)$, $\mathcal{S}(v_n^z)$ and other velocity cross correlation terms. For $v_{ns}=v_{ns}^{(2)}=0.94~\rm cm/s$ (as experiments in 
Ref.\cite{svancara-etal-2021} are performed at large $v_{ns}$) we obtain the following values for the skewnesses (in non-dimensional units scaled
with unit of length $\lambda=1.59\times 10^{-2}$cm and unit of time $\tau=1.22 \times 10^{-2}$s): $\mathcal{S}(v_L^z)=-2.3\times 10^{-4}$, 
$\mathcal{S}(v_{_{BS}}^z)=-1.8\times 10^{-4}$ and $\mathcal{S}(v_n^z)=-8\times 10^{-4}$. Hence, the skewness of $v_{_{BS}}^z$ does not fully account for the
skewness of $v_L^z$, implying that the skewness of $v_n^z$ arising from the existence of the wakes, contributes to $\mathcal{S}(v_L^z)$ and hence to the non symmetrical character of the statistical distributions of the velocity of inertial particles observed in experiments \cite{svancara-etal-2021}. As
$\varepsilon_2$ increases for increasing temperature, the impact of the wakes on the experimental particle velocity statistics is likely to increase with temperature.

\end{document}